\RequirePackage{fix-cm}
\documentclass[smallextended]{svjour3}       
\smartqed  
\usepackage{graphicx}
\usepackage[]{hyperref}
\usepackage{booktabs}
\usepackage{multirow}
\usepackage{amssymb}
\newcommand{\orcid}[1]{\href{https://orcid.org/#1}{\includegraphics[scale=0.25]{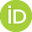}}}
\begin{document}

\title{Fully-automated Body Composition Analysis in Routine CT Imaging Using 3D Semantic Segmentation Convolutional Neural Networks
}

\titlerunning{Fully-automated Body Composition Analysis}        

\author{Sven Koitka \orcid{0000-0001-9704-1180} \and
		Lennard Kroll \orcid{0000-0001-8102-6146} \and
		Eugen Malamutmann \orcid{0000-0003-4624-5890} \and \\
		Arzu Oezcelik \orcid{0000-0002-8353-9532} \and
		Felix Nensa \orcid{0000-0002-5811-7100}
}

\authorrunning{Koitka et al.} 

\institute{Sven Koitka \and Lennard Kroll \and Felix Nensa \at
           University Hospital Essen, Institute of Diagnostic and Interventional Radiology and Neuroradiology, Essen, Germany \\
              \email{sven.koitka@uk-essen.de}           
           \and
           Eugen Malamutmann \and Arzu Oezcelik \at
           University Hospital Essen, Department of General, Visceral, and Transplantation Surgery, Essen, Germany          
}

\date{Received: - / Accepted: -}

\maketitle

\begin{abstract}
Body tissue composition is a long-known biomarker with high diagnostic and prognostic value in cardiovascular, oncological and orthopaedic diseases, but also in rehabilitation medicine or drug dosage. In this study, the aim was to develop a fully automated, reproducible and quantitative 3D volumetry of body tissue composition from standard CT examinations of the abdomen in order to be able to offer such valuable biomarkers as part of routine clinical imaging. Therefore an in-house dataset of 40 CTs for training and 10 CTs for testing were fully annotated on every fifth axial slice with five different semantic body regions: abdominal cavity, bones, muscle, subcutaneous tissue, and thoracic cavity. Multi-resolution U-Net 3D neural networks were employed for segmenting these body regions, followed by subclassifying adipose tissue and muscle using known hounsfield unit limits. The S{\o}rensen Dice scores averaged over all semantic regions was 0.9553 and the intra-class correlation coefficients for subclassified tissues were above 0.99. Our results show that fully-automated body composition analysis on routine CT imaging can provide stable biomarkers across the whole abdomen and not just on L3 slices, which is historically the reference location for analysing body composition in the clinical routine.
\keywords{Abdomen \and Body Composition \and Computer Tomography \and Deep Learning \and Semantic Segmentation}
\pagebreak
\noindent\textbf{Keypoints}
\begin{itemize}
	\item Our study enables fully automated body composition analysis on routine abdomen CT scans
	\item The best segmentation models for semantic body region segmentation achieved an averaged S{\o}rensen Dice score of 0.9553
	\item Subclassified tissue volumes achieved intra-class correlation coefficients over 0.99
\end{itemize}
\end{abstract}

\section{Introduction}
\label{intro}
Thanks to advances in computer-aided image analysis, radiological image data are now increasingly considered a valuable source of quantitative biomarkers. Body tissue composition is a long-known biomarker with high diagnostic and prognostic value in cardiovascular, oncological and orthopaedic diseases, but also in rehabilitation medicine or drug dosage. As obvious and simple as a quantitative determination of tissue composition based on modern radiological sectional imaging may seem, the actual extraction of this information in clinical routine is not feasible, since a manual assessment requires an extraordinary amount of human labour. A recent study has shown that some anthropometric measures can be estimated from simple and reproducible 2D measurements in CT using linear regression models \cite{Zopfs2019}. Another study showed that a fully automated 2D segmentation of CT sectional images at the level of L3 vertebra into subcutaneous adipose tissue, muscle, viscera, and bone was possible using a 2D U-Net architecture \cite{Weston2019}. The determination of the tissue composition at the level of L3 is often used as a reference in clinical routine to limit the amount of work required for the assessment. However, even here this is only a rough approximation, since the inter-individual variability between patients is large and the section at the level of L3 does not necessarily have to be representative of the entire human anatomy. Other dedicated techniques for analyzing body composition using Dual‐energy X‐ray absorptiometry or Magnetic resonance imaging exist \cite{Seabolt2015} but require additional potentially time-consuming or expensive procedures to be performed. 

The aim of our study was therefore to develop a fully automated, reproducible and quantitative 3D volumetry of body tissue composition from standard CT examinations of the abdomen in order to be able to offer such valuable biomarkers as part of routine clinical imaging.

\section{Materials and Methods}

\subsection{Dataset}

\begin{figure}[htbp]
	\includegraphics[width=\linewidth]{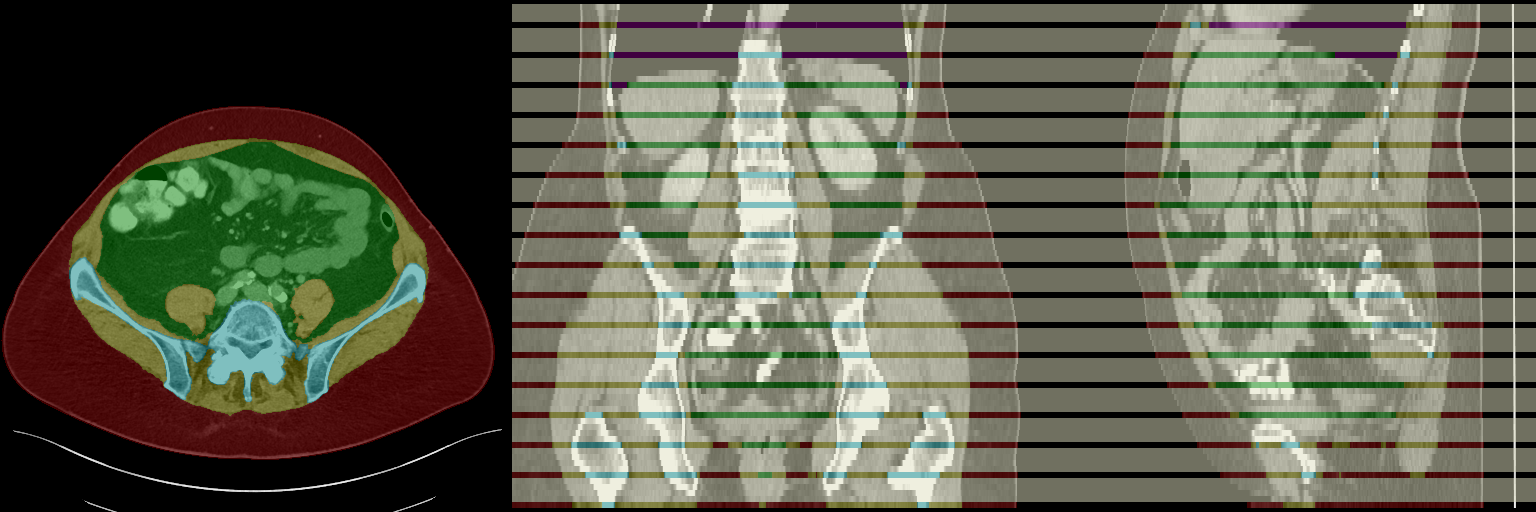}
	\caption{Exemplary annotation of an abdominal CT, with subcutaneous tissue (red), muscle (yellow), bones (blue), abdominal cavity (green), thoracic cavity (purple), and ignore regions (white).}
	\label{fig:annotation}
\end{figure}

A retrospective dataset was collected, consisting of 40 abdominal CTs for training and 10 abdominal CTs for testing. The included scans were randomly selected from abdominal CT studies performed between 2015 and 2019 at the University Hospital Essen, Germany. Each CT volume has a slice thickness of 5mm and was reconstructed using a soft tissue convolutional reconstruction kernel. The data was annotated with five different labels: background (= outside the human body), muscle, bones, subcutaneous tissue, abdominal cavity, and thoracic cavity. For annotation, the ITK Snap \cite{Yushkevich2006} software (version 3.8.0) was used. Region segmentation was performed manually with a polygon tool. In order to reduce the annotation effort, every fifth slice was fully annotated. Remaining slices were marked with an ignore label, as visualized in Figure \ref{fig:annotation}. The final dataset contains 751 fully annotated slices for training and 186 for testing.

\subsection{Network Architectures}
\label{sec:architecture}

\begin{figure}[htbp]
	\includegraphics[width=\linewidth]{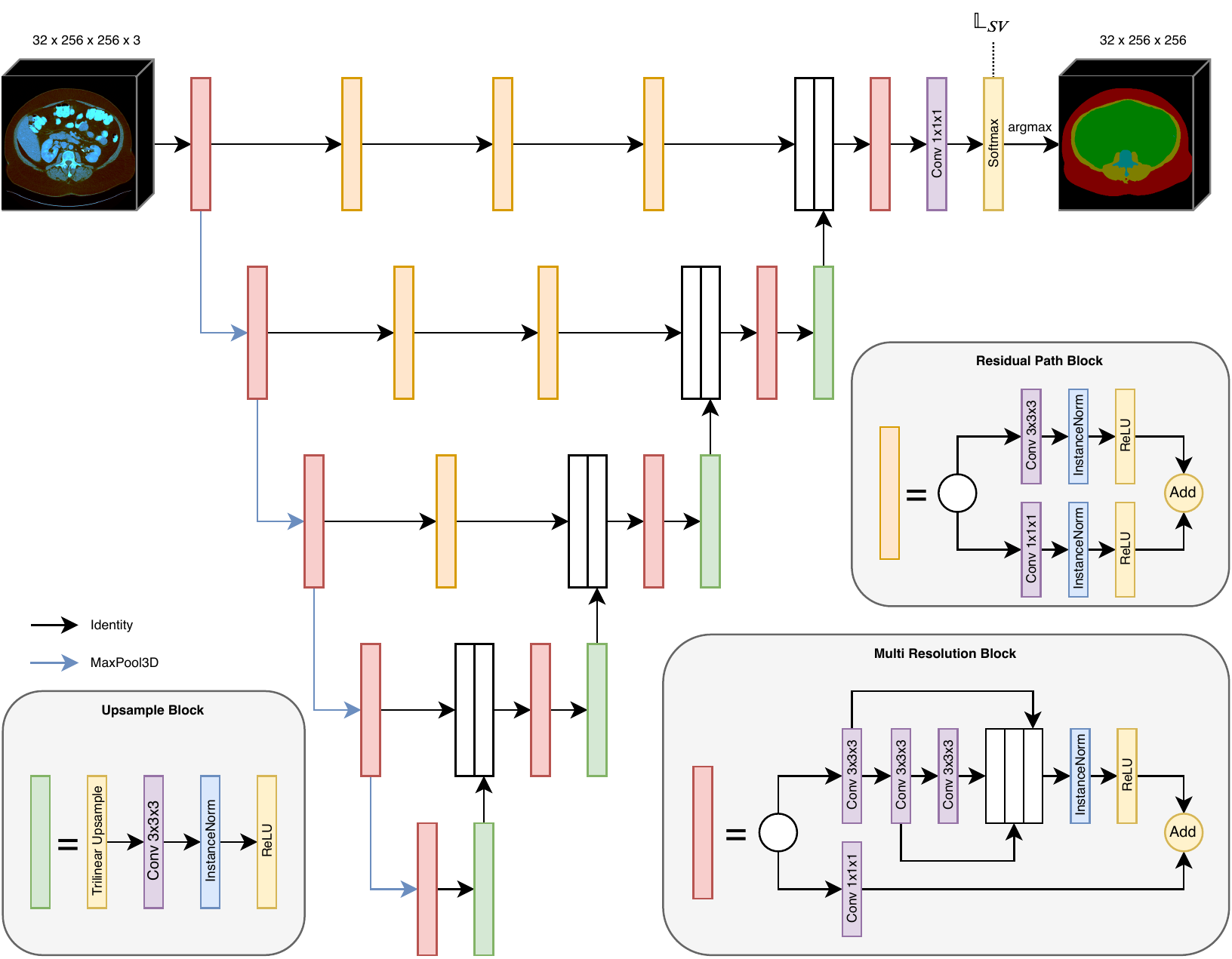}
	\caption{Schematic overview of the multi-resolution U-Net 3D architecture.}
	\label{fig:network}
\end{figure}

For this study two different network architectures were chosen for training, namely the U-Net 3D \cite{Cicek2016} and a more recent variant multi-resolution U-Net 3D \cite{Ibtehaz2020}. The latter is shown in Figure \ref{fig:network}, however, U-Net 3D is very similar with residual path blocks replaced by identity operations and multi resolution blocks replaced by two successive convolutions. In this case, volumetric data limits the batch size to a single example per batch due to a large memory footprint. Therefore, instance normalization layers \cite{Ulyanov2017} were utilized in favor of batch normalization layers \cite{Ioffe2015}. In the original architectures, transposed convolutions were employed to upsample feature maps back to the original image size. However, transposed convolutions tend to generate checkerboard artifacts \cite{odena2016}. This is why trilinear upsampling followed by a  convolution was used instead, which is computationally more expensive, but more stable during optimization. Additionally, different choices for the initial number of feature maps are evaluated: 16, 32, and 64. After each pooling step the number gets doubled, resulting in 256, 512, and 1024 feature maps in the lowest resolution, respectively.

\subsection{Training Details}
\label{sec:training}

The implementation of network architectures and training was done using Tensorflow 2.0 \cite{Abadi2016} and the Keras API. Nvidia Titan RTX GPUs with 24GB VRAM were used, which enable training of more complex network architectures when using large volumetric data.

Adam \cite{Kingma2015} with decoupled weight decay regularization \cite{Loshchilov2019} was utilized, configured with $beta_1$=0.9, $beta_2$=0.999, eps=1e-7, and weight decay of 1e-4. An exponentially decaying learning rate with initial value of 1e-4, multiplied by 0.95 every 50 epochs, helped to stabilize the optimization process at the end of the training. For selecting the best model weights during training, 5-fold cross-validation was used on the training set and the average dice score was monitored on the respective validation splits. Since the training dataset consists of 40 abdominal CTs, each training run was performed using 32 CTs for training and 8 CTs for validation.

During training, several data augmentations were applied in order to virtually increase the unique sample size for training a generalizable network. First, random scale augmentation was applied with a scaling factor sampled uniformly between 0.8 and 1.2. Since this factor was sampled independently for both x- and y axis, it also acts as an aspect ratio augmentation. Second, random flipping was utilized to mirror volumes on the x-axis. Third, subvolumes of size $32\times256\times256$ were randomly cropped from the full volume with size $n\times512\times512$. During inference, the same number of slices was used, but with x and y dimension kept unchanged, and the whole volume was processed using a sliding window approach with 75\% overlap. To improve segmentation accuracy, predictions for overlapping subvolumes were aggregated in a weighted fashion, giving the central slices more weight than the outermost.

Besides random data augmentations, additional pre-processing steps were performed before feeding the image data into the neural networks. Volumes were downscaled by factor 2 to  on the x/y axes, retaining a slice thickness of 5mm on the z-axis. CT images are captured as hounsfield units (HU), which capture fine details and allow for different interpretations depending on which transfer function is used to map HUs to a color (e.g. black/white). Normally, when using floating point values the typical scanner quantization of 12 bits can be stored lossless and a network should be able to process all information without any problems. In this work, multiple HU windows [-1024, 4096], [-150, 250],  and [-95, 155] were applied with clipping outliers to the respective minimum and maximum values and stacked as channels. Lastly, the network inputs were centered around zero with minimum value at -1 and maximum value at +1.

For supervision, a combination of softmax cross entropy loss and generalized S{\o}rensen Dice loss \cite{Sudre2017} was chosen, similar to \cite{Isensee2019}. Both losses are defined as below:
\begin{equation}
	\mathbb{L}_{XCE} = -\frac{1}{N} \cdot \sum_{n=1}^{N} \sum_{c=1}^{C} y_{c,n} \cdot \log\left(\hat{y}_{c,n}\right)
\end{equation}
\begin{equation}
	\mathbb{L}_{Dice} = 1.0 - \frac{1}{C-1}\cdot\sum_{c=2}^{C}\frac{\sum_{n=1}^{N}2\cdot\hat{y}_{c,n}\cdot y_{c,n} + \epsilon}{\sum_{n=1}^{N}\hat{y}_{c,n} + y_{c,n} + \epsilon}
\end{equation}
$C$ stands for for the total number of classes, which equals six for the problem at hand. $\hat{y}_{c,n}$ and $y_{c,n}$ represent the prediction respectively groundtruth label for class $c$ at voxel location $n$. The background class is in this work explicitly not covered by the dice loss in order to give the foreground classes more weight in the optimization process. This choice is well known for class imbalanced problems where the foreground class only covers little areas compared to the background class.

The final loss is an equally weighted combination of both losses:
\begin{equation}
	\mathbb{L}_{SV} = 0.5 \cdot \mathbb{L}_{XCE} + 0.5 \cdot \mathbb{L}_{Dice}
\end{equation}

\subsection{Tissue Quantification}
\label{sec:method:report}

Various materials can be extracted from a CT by thresholding the HU to a specific intensity range. For quantifying tissues, the reporting system uses a mixture of classical thresholding and modern semantic segmentation neural networks for building the semantic relationships. During training, five models were optimized using cross-validation to evaluate the generalization performance. When using these for inference, the predictions of all five models are averaged and thus an ensemble model is constructed. The final output of the quantification system is a report about subcutaneous adipose tissue (SAT), visceral adipose tissue (VAT), and muscle volume. Muscular tissue is identified by thresholding the HU between -29 and 150. Adipose tissue is identified by thresholding the HU between -190 and -30. If an adipose voxel is within the abdominal cavity region, it is counted as VAT. If it is within the subcutaneous tissue region, it is counted as SAT.

\section{Results}

\subsection{Model Evaluation}

As described in Section \ref{sec:architecture} and \ref{sec:training}, two different network architectures with varying initial number of feature maps were systematically evaluated using a 5-fold cross-validation scheme on the training dataset. The results are stated in Table 1. First of all, all networks delivered promising results with average dice scores over 0.93. Second, multi-resolution U-Net variants achieved constantly higher scores compared to their respective U-Net counterparts. It is interesting to note, that the improvements in scores were small compared to the increase in trainable parameters and thus required time to train and test the networks. A single optimization step took 294ms, 500ms, and 1043ms on a NVIDIA Titan RTX for the initial feature map count of 16, 32, and 64, respectively.

\begin{table}[htbp]
	\small
	\setlength{\tabcolsep}{3pt}
	\renewcommand{\arraystretch}{1.5}
	\caption{Evaluation for 5-fold cross-validation runs (stated as mean over all runs) and ensemble predictions on the test set. (AC) Abdominal Cavity, (B) Bones, (M) Muscle, (ST) Subcutaneous Tissue, (TC) Thoracic Cavity.}
	\begin{center}
	\begin{tabular}{llcrcccccc}
		\toprule
		& & & & \multicolumn{5}{c}{\textbf{Dice Score}}\\
		\cline{5-9}
		&\textbf{Model} & $n_f$ & $n_{param}$ & AC & B & M & ST & TC & Average \\
		\midrule
		\multirow{6}{*}{\rotatebox[origin=c]{90}{\textbf{5-fold CV}}} \quad & U-Net 3D & 16 & 5.34M & $0.9509$ & $0.9462$ & $0.9266$ & $0.9432$ & $0.8823$ & $0.9299$\\
		& & 32 & 21.36M & $0.9669$ & $0.9540$ & $0.9379$ & $0.9574$ & $0.9336$ & $0.9500$\\		
		& & 64 & 85.43M & $0.9682$ & $0.9561$ & $0.9403$ & $0.9582$ & $0.9481$ & $0.9542$\\
		\cline{2-10}
		& multi-res U-Net 3D & 16 & 5.82M & $0.9589$ & $0.9484$ & $0.9328$ & $0.9531$ & $0.9211$ & $0.9429$ \\
		& & 32 & 21.24M & $0.9680$ & $0.9554$ & $0.9399$ & $0.9596$ & $0.9414$ & $0.9529$ \\		
		& & 64 & 85.10M & $0.9692$ & $0.9564$ & $0.9414$ & $0.9605$ & $0.9452$ & $0.9545$\\
		\midrule
		\multirow{6}{*}{\rotatebox[origin=c]{90}{\textbf{Test Set}}}& U-Net 3D & 16 & 5.34M & $0.9609$ & $0.9340$ & $0.9229$ & $0.9553$ & $0.9172$ & $0.9381$\\
		& & 32 & 21.36M & $0.9731$ & $0.9390$ & $0.9309$ & $0.9610$ & $0.9598$ & $0.9528$ \\		
		& & 64 & 85.43M & $0.9739$ & $0.9406$ & $0.9316$ & $0.9623$ & $0.9641$ & $0.9545$\\		
		\cline{2-10}
		& multi-res U-Net 3D & 16 & 5.82M & $0.9667$ & $0.9355$ & $0.9272$ & $0.9593$ & $0.9518$ & $0.9481$ \\
		& & 32 & 21.24M & $0.9736$ & $0.9409$ & $0.9328$ & $0.9627$ & $0.9629$ & $0.9546$ \\		
		& & 64 & 85.10M & $0.9735$ & $0.9423$ & $0.9334$ & $0.9623$ & $0.9652$ & $0.9553$\\
		\bottomrule
	\end{tabular}
	\end{center}
	\label{tab:model-evaluation}
\end{table}

\begin{figure}[htbp]
	\includegraphics[width=\linewidth]{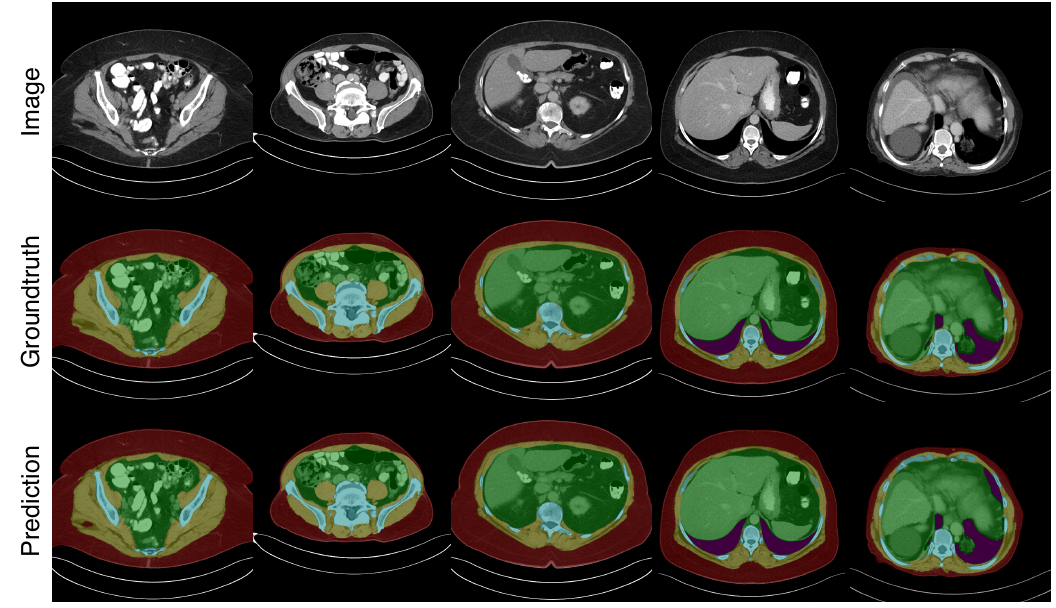}
	\caption{Comparison of different slices, their respective groundtruth annotation and predictions of the ensemble formed from five trained models on cross-validation splits.}
	\label{fig:prediction-vs-groundtruth}
\end{figure}

For visual inspection of the ensemble segmentations, a few exemplary slices are shown in Figure \ref{fig:prediction-vs-groundtruth}. Most slices show almost perfect segmentation boundaries, however, especially the ribs are problematic due to the partial volume effect. In 5mm CTs it is even for human readers sometimes hard to correctly assign one or the other region.

\subsection{Ablation Study}

During model development it was observed, that the choice of HU window has an impact on optimization stability and final achieved scores. Therefore a small ablation study was conducted in order to systematically evaluate the influence of different HU limits. Additional models were trained using the same training parameters, but only with changed input pre-processing. The results are stated in Table \ref{tab:ablation}.

\begin{table}[htbp]
	\small
	\setlength{\tabcolsep}{3pt}
	\renewcommand{\arraystretch}{1.5}
	\caption{Evaluation of multi-resolution U-Nets with $n_f = 32$ trained on different mappings from hounsfield units to the target intensity value range of $[-1, 1]$. Multi-Window stands for a combination of theoretical value range of 12-bit CT scans, abdomen window, and liver window. (AC) Abdominal Cavity, (B) Bones, (M) Muscle, (ST) Subcutaneous Tissue, (TC) Thoracic Cavity.}
	\begin{center}
	\begin{tabular}{llcccccc}
		\toprule
		& & \multicolumn{5}{c}{\textbf{Dice Score}}\\
		\cline{3-7}
		&\textbf{HU Window} & AC & B & M & ST & TC & Average \\
		\midrule
		\multirow{5}{*}{\rotatebox[origin=c]{90}{\textbf{5-fold CV}}} \quad &Multi-Window & $0.9680$ & $0.9554$ & $0.9399$ & $0.9596$ & $0.9414$ & $0.9529$ \\		
		&$[-1024,3071]$ & $0.9561$ & $0.9403$ & $0.9217$ & $0.9494$ & $0.9254$ & $0.9386$ \\		
		&$[-1024,2047]$ & $0.9533$ & $0.9410$ & $0.9144$ & $0.9412$ & $0.9303$ & $0.9360$ \\		
		&$[-1024,1023]$ & $0.8731$ & $0.8778$ & $0.7875$ & $0.6959$ & $0.8696$ & $0.8208$ \\		
		&$[-150,250]$ & $0.8598$ & $0.8687$ & $0.7632$ & $0.7772$ & $0.8759$ & $0.8289$ \\
		
		\midrule
		
		\multirow{5}{*}{\rotatebox[origin=c]{90}{\textbf{Test Set}}}&Multi-Window & $0.9736$ & $0.9409$ & $0.9328$ & $0.9627$ & $0.9629$ & $0.9546$ \\
		&$[-1024,3071]$ & $0.9682$ & $0.9392$ & $0.9261$ & $0.9606$ & $0.9532$ & $0.9495$ \\
		&$[-1024,2047]$ & $0.9644$ & $0.9331$ & $0.9174$ & $0.9560$ & $0.9569$ & $0.9455$ \\
		&$[-1024,1023]$ & $0.9329$ & $0.9002$ & $0.8412$ & $0.8879$ & $0.9066$ & $0.8938$ \\
		&$[-150,250]$ & $0.8950$ & $0.8997$ & $0.8004$ & $0.8482$ & $0.9311$ & $0.8749$ \\
		\bottomrule
	\end{tabular}
	\end{center}
	\label{tab:ablation}
\end{table}

Increasing the HU intensity range consistently improves dice scores. By combining multiple HU windows as separate input channels, the dice scores can be even more improved to over 0.95 dice score on average on both cross-validation and test set. The lowest scores of 0.829 dice on average for cross-validation and 0.875 for the test set were achieved by an abdominal HU window ranging from -150 to 250.

\subsection{Tissue Quantification Report}

\begin{figure}[htbp]
	\includegraphics[width=\linewidth]{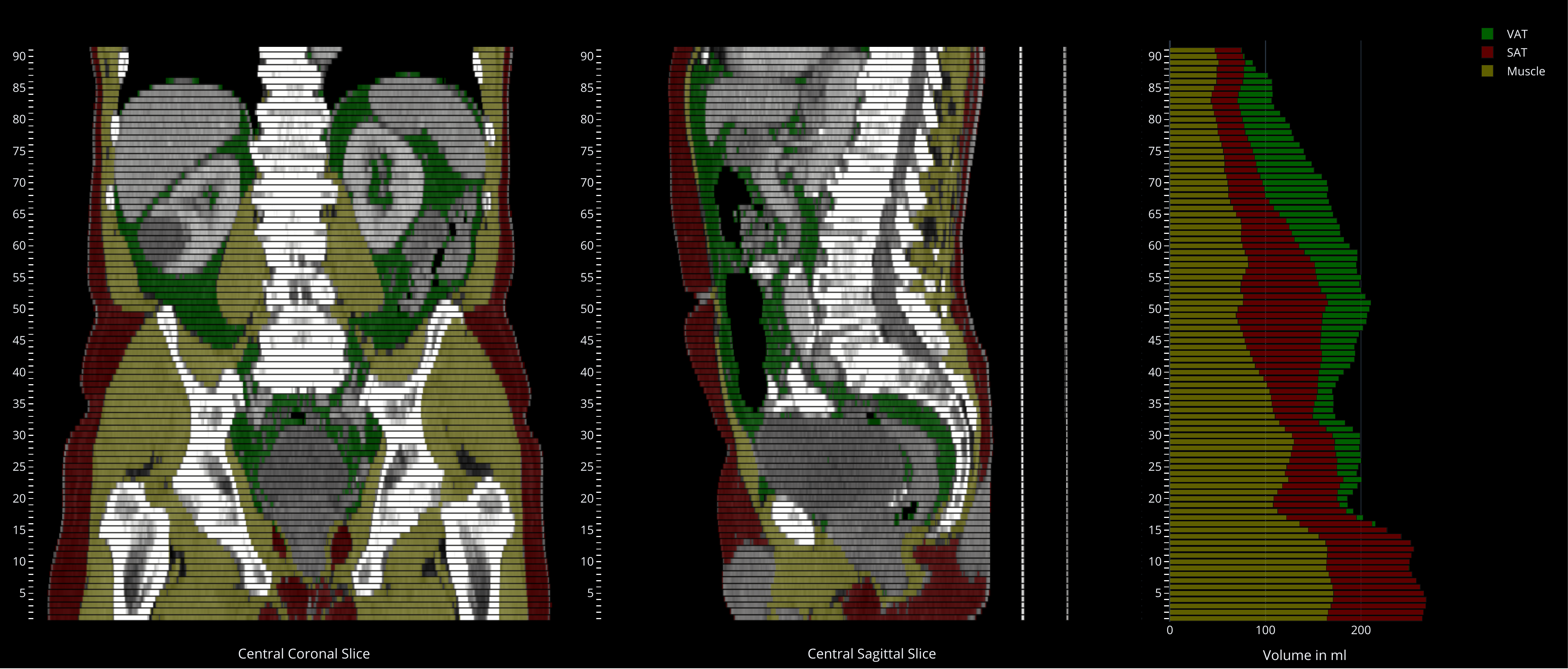}
	\caption{Final visual report of the tissue quantification system output. SAT is shown in red, VAT is shown in green, and muscle tissue is shown yellow.}
	\label{fig:report}
\end{figure}

As described in Section \ref{sec:method:report}, the segmentation models are intended to be used for assigning thresholded tissues to different regions, which is technically a logical conjunction. The achieved intraclass correlation coefficients for the derived SAT, VAT, and muscle volumes measured per slice are 0.999, 0.998, and 0.991, respectively ($p<0.001$). In order to visually inspect the quality of the tissue segmentation, a PDF report with sagittal and coronal slices is generated, in conjunction with a stacked bar plot showing the volumes of segmented muscle, SAT, and VAT per axial slice (see Figure \ref{fig:report}). This is only intended to give the human reader a first visual impression on the system output. For analysis, an additional table with all numeric values per slice is generated. The PDF file is encapsulated into DICOM and automatically sent back to the PACS, in order to make use of existing DICOM infrastructure.

\section{Discussion}

Our study aimed to develop a fully automated, reproducible and quantitative 3D volumetry of body tissue composition from standard abdominal CT examinations in order to provide valuable biomarkers as part of routine clinical imaging.

Our best approach using a multi-resolution U-Net 3D with an initial feature map count of 64 was able to fully automatically segment abdominal cavity, bones, muscle, subcutaneous tissue, and thoracic cavity with a mean S{\o}rensen Dice coefficient of 0.9553 and thus yielded excellent results. The derived tissue volumetry had intraclass correlation coefficients of over 0.99. Further experiments showed a high performance with heavily reduced parameter counts which enables considering speed / accuracy trade-offs depending on the type of application. Choosing the transfer function to map from HU to a normalized value range for feeding images into neural networks was found to have a huge impact on segmentation performance.

In a recent study, manual single-slice CT measurements were used to build linear regression models for predicting stable anthropometric measures \cite{Zopfs2019}. As the authors suggest, these measures may be important as biomarkers for several diseases like e.g. sarcopenia, but could also be used where the real measurements are not available. However, manual single-slice CT measurements are still prone to intra-patient variability and inter- as well as intra-rater variability. By using a fully automated approach, derived anthropometric measures from more than a single CT slice should in theory be more stable.

Fully automated analysis of body composition has been attempted many times in the past. Older methods utilize classical image processing and binary morphological operations \cite{Kim2013,Kullberg2017,Mensink2011} in order to isolate SAT and VAT from total adipose tissue (TAT). Other studies use prior knowledge about contours and shapes and actively fit a contour or template to a given CT image \cite{Agarwal2017,Ohshima2008,Parikh2017,Pednekar2005,Popuri2016}. Those methods are prone to variations in intensity values and assume certain body structures for algorithmic separation between SAT and VAT.  Apart from purely CT imaging based studies there have been efforts to apply similar techniques to magnetic resonance imaging (MRI) \cite{Joshi2013,Positano2004,Zhou2011}. However, MRI procedures are more cost and time expensive than CT imaging in the clinical routine. Specific MRI procedures exist for body fat assessment, but have to be performed explicitly. Our approach can be used on routine CT imaging and may be used as supplementary material for diagnosis or screening purposes.

Recently, deep learning based methods have been proposed \cite{Bridge2018,Weston2019}. In both studies, models were trained solely on single L3 CT slices. However, Weston et al. \cite{Weston2019} visually showed that their model was able to generalize for other abdominal slices well without being trained on such data. Nonetheless, they mentioned that extending the training and evaluation data to the whole abdomen would be beneficial for stability but also analysis capabilities. Our study uses annotated data for training and evaluation across the whole abdomen and thus is a true volumetric approach to body composition analysis. In addition, they segmented SAT and VAT directly, whereas in our study the semantic body region was segmented and adipose tissue was subclassified using known HU thresholds.

\begin{figure}[b]
	\includegraphics[width=\linewidth]{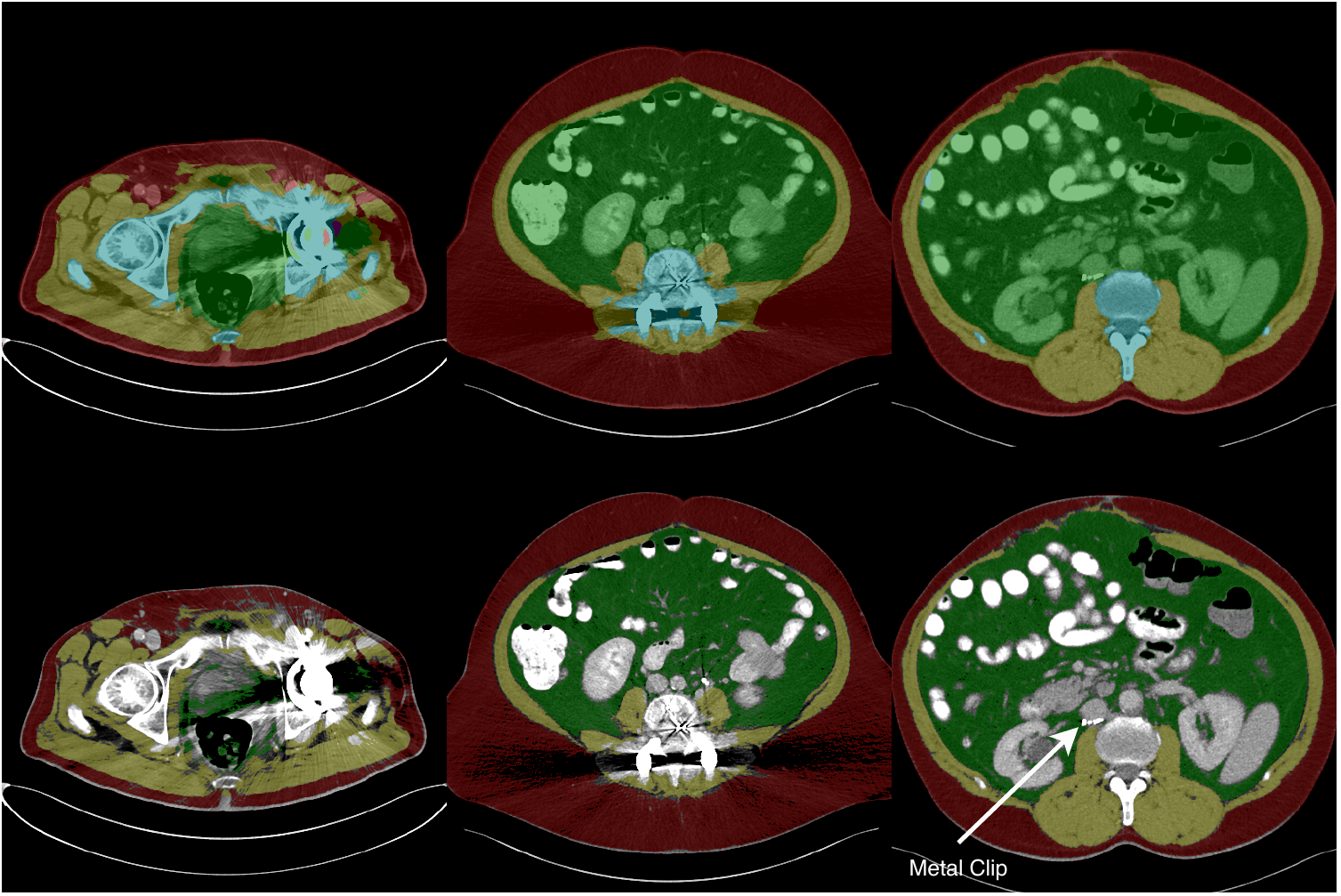}
	\caption{Beam hardening artifacts may not only harm segmentation quality (top), but also prevent accurate identification of tissues (bottom). (left) Strong beam hardening artifacts with faults in the segmentation output (middle) Beam hardening artifacts with mostly accurate segmentation, but streaking artifacts prevent accurate muscle and SAT identification (right) No beam hardening artifacts at all, but metal foreign object detected.}
	\label{fig:artifacts}
\end{figure}

One major disadvantage of the collected dataset is the slice thickness of 5mm. Several tissues, materials, and potentially air can be contained within a distance of 5mm, the resulting HU at a specific location is an average of all components. This is also known as partial volume effect and can be counteracted by using a smaller slice thickness, ideally with isometric voxel sizes. However, a reconstructed slice thickness of 5 mm is common in clinical routine CT and it is questionable whether the increased precision of calculating the tissue composition on 1 mm slices would have clinical relevance. Nevertheless, we plan to investigate the influence of thinner slices in further studies, as the reading on thin slices is becoming routine in more and more institutions.  

Another limitation is the differentiation between visceral fat and fat contained within organs. Currently, every voxel with HU in the fat intensity value range, which is contained within the abdominal cavity region, is counted as VAT. However, per definition, fat cells within organs do not count as VAT and thus should be excluded from the final statistics. Public datasets like \cite{Gibson2018,Gibson2018b} already exist for multi-organ semantic segmentation and could be utilized to postprocess the segmentation results from this study by masking organs in the abdominal cavity.

It is quite common to find metal foreign objects like implants in abdominal CTs and thus to encounter beam hardening artifacts. Those artifacts, depending on how strong they are, may affect the segmentation quality, as shown in Figure \ref{fig:artifacts}. Even if the segmentation model is able to predict the precise boundary of the individual semantic regions, streaking and cupping artifacts make it impossible to threshold fatty or muscular tissue based on HU intensities potentially invalidating quantification reports. In a future version of our tool we are therefore planning a functionality for automatic detection and handling of image artifacts.

In future works, we plan to extend the body composition analysis system to incorporate other regions of the body as well. For example, \cite{Kullberg2017} already showed an analysis of adipose tissue and muscle for thighs. Ideally, the system should be capable of analysing the whole body in order to derive stable biomarkers.

\section{Conclusion}

In the present study we presented a deep learning based, fully automated volumetric tissue classification system for the extraction of robust biomarkers from clinical CT examinations of the abdomen. In the future, we plan to extend the system to thoracic examinations and to add important tissue classes such as pericardial adipose tissue and myocardium.


%



%
%

\end{document}